# A photon-counting detector for dual-energy breast tomosynthesis

Erik Fredenberg,[a] Mats Lundqvist,[b] Magnus Åslund,[b] Magnus Hemmendorff,[b] Björn Cederström,[a] and Mats Danielsson[a]

[a]Department of Physics, Royal Institute of Technology (KTH), AlbaNova University Center, SE-106 91 Stockholm, Sweden

[b]Sectra Mamea AB, Smidesvägen 5, SE-171 41 Solna, Sweden;

**ABSTRACT**

We present the first evaluation of a recently developed silicon-strip detector for photon-counting dual-energy breast tomosynthesis. The detector is well suited for tomosynthesis with high dose efficiency and intrinsic scatter rejection. A method was developed for measuring the spatial resolution of a system based on the detector in terms of the three-dimensional modulation transfer function (MTF). The measurements agreed well with theoretical expectations, and it was seen that depth resolution was won at the cost of a slightly decreased lateral resolution. This may be a justifiable trade-off as clinical images acquired with the system indicate improved conspicuity of breast lesions. The photon-counting detector enables dual-energy subtraction imaging with electronic spectrum-splitting. This improved the detectability of iodine in phantom measurements, and the detector was found to be stable over typical clinical acquisition times. A model of the energy resolution showed that further improvements are within reach by optimization of the detector.

## 1. INTRODUCTION

Breast tomosynthesis[1–3] and contrast-enhanced dual-energy subtraction (DES) imaging,[4–7] are two promising methods to improve the conspicuity of breast lesions in mammography, offering an alternative to MRI. A photon-counting scanned multi-slit mammography system with an array of silicon-strip detectors has recently been adapted for breast tomosynthesis within the EU-funded HighReX-project.[8] Clinical trials have begun during the fall.

The HighReX system is particularly well suited for tomosynthesis for at least two reasons. (1) Multi-slit geometries provide intrinsic and efficient scatter rejection.[9] This is contrary to systems that rely on Bucky grids, which are relatively inefficient for mammography energies and hard to integrate in tomosynthesis. Scatter rejection is important in tomosynthesis,[2] in particular for large and dense breasts where tomosynthesis is expected to be most beneficial because of the reduced conspicuity of lesions. (2) Compared to energy-integration, photon counting provides improved energy weighting,[10] and it is possible to eliminate virtually all electronic noise.[11] The result is higher dose efficiency, which is important in tomosynthesis since the dose in each projection must be kept low.

Moreover, photon counting allows the energy of individual photons to be measured.[12] The silicon-strip detector that is used in the described tomosynthesis system is able to separate impinging photons into two bins according to their energy. One application of such energy sensitivity is contrast-enhanced DES imaging using electronic spectrum-splitting, which is more efficient than dual-spectra methods because of reduced motion artifacts and narrower spectra.[6,7] The detector is also equipped with anti-coincidence logic to reduce the effects of charge sharing between adjacent strips and improve the energy resolution.

We present an evaluation of the HighReX system, divided into two parts; tomosynthesis and DES imaging. Tomosynthesis was studied on a system level in a prototype. Clinical images have been acquired, and the system

---

Electronic mail: fberg@mi.physics.kth.se

was characterized in terms of the resolution. DES is not yet implemented on this first system, and energy sensitivity was examined on a detector level using a single detector line. We demonstrate a model of the energy characteristics of the detector, which is based on theoretical considerations and measurements.

## 2. MATERIAL AND METHODS

### 2.1. Background

#### 2.1.1. Tomosynthesis

Tomosynthesis allows for three-dimensional reconstruction from a set of limited-angle projections. Throughout the study, the $x$-$y$-plane refers to the lateral directions in the object, with detector strips in the $x$-direction and tube motion and scanning along $y$. The $z$-axis is the depth direction. $\nu_x$, $\nu_y$, and $\nu_z$ are the respective spatial frequencies.

Tomosynthesis reconstruction can be approximately described with the Fourier-slice theorem, stating that the Fourier transform of a projection line is equal to a slice at the same angle through the origin of the two-dimensional (2D) Fourier transform of the reconstructed plane.[3] As opposed to computed tomography with a full angular coverage, tomosynthesis restrains the Fourier domain of the reconstructed plane to a sheaf-shaped area around the origin. One implication of the restricted Fourier plane is a limited $z$-resolution, in particular for low $\nu_y$, which show up as shadows that transfer through the planes. Another implication is an edge enhancement in $y$.

There is not yet a standardized method of measuring tomosynthesis resolution. The $y$-resolution may be measured separately from $z$ with standard methods, but the results can be misleading since the influence of out-of-plane structures is not taken into account. Instead, we make an attempt to estimate the 2D modulation transfer function (MTF) in the $y$-$z$-plane, based on a measurement of the 2D point-spread function (PSF). In $x$, the resolution is virtually unaffected by the reconstruction process and determined by the source size and the detector. Being independent of $y$ and $z$, the $x$-MTF can be calculated from a measurement of the one-dimensional (1D) line-spread function.

#### 2.1.2. Contrast-enhanced DES imaging with electronic spectrum-splitting

DES imaging with electronic spectrum-splitting as used in the HighReX system has been described in detail previously,[6,7] but is reviewed here for clarity. An x-ray detector with an energy threshold makes it possible to simultaneously record high- and low-energy images ($n_{\text{hi}}$ and $n_{\text{lo}}$) by sorting registered photons into two bins. By logarithmically subtracting the two images with a proper weighting factor ($w$) the contrast between any two materials (e.g. glandular and adipose tissue) in the object can be made to cancel, whereas other materials are still visible. The effect is largest if the spectrum is split at an absorption edge of a contrast agent material. The DES signal ($I_{\text{DES}}$) and signal difference ($\Delta I_{\text{DES}}$) are thus formed;

$$I_{\text{DES}} = \ln n_{\text{hi}} - w \ln n_{\text{lo}} \qquad \text{and} \qquad \Delta I_{\text{DES}} = \left| I_{\text{DES}}^1 - I_{\text{DES}}^2 \right|. \tag{1}$$

It has been shown that $\Delta I_{\text{DES}}$ is independent of exposure.[7]

$w$ is chosen so as to minimize the noise from the anatomical background, which is calculated as the standard deviation of $I_{\text{DES}}$ measured with negligible quantum noise over a range of glandular fractions, $g_i$;

$$\sigma_{\text{w}}(w) = \left( \frac{1}{m-1} \sum_i^m \left[ I_{\text{DES}}(g_i, w) - \overline{I_{\text{DES}}(w)} \right]^2 \right)^{1/2}, \tag{2}$$

where $\overline{I_{\text{DES}}(w)}$ is the mean over all glandular fractions. Note that $w$ is also exposure-independent.

## 2.2. Description of the system and experimental set-up

### 2.2.1. The tomosynthesis system

The HighReX tomosynthesis system is a modification of the Sectra MicroDose Mammography (MDM) system.[9, 13] The MDM system is a scanning multi-slit full-field digital mammography system, which comprises a mammography x-ray tube, a pre-collimator, and an image receptor, all mounted on a common arm (Fig. 1, left). To acquire an image, the arm is rotated around the center of the source so that the detector and pre-collimator are scanned across the object. In the HighReX system, however, the center of rotation is shifted to a point below the detector so that the source movement describes an arc above the object, while the detector and pre-collimator scan the object. Hence, each point in the object is viewed from a range of source-detector angles by different detector segments as the detector scans the object. The angular coverage is limited by the width of the detector to 11°. This scheme is different from tomosynthesis with flat-panel detectors as two-dimensional projection images are not acquired successively, but the reconstruction principle is equal. The advantages include more efficient scatter rejection and less focal spot blurring as the detector moves with the source. Figure 1, right shows a photograph of the HighReX system. The source and beam quality are similar to the MDM system.

The image reconstruction is based on the convex algorithm introduced for transmission imaging by Lange.[14] It is an iterative method, similar to expectation maximization, with the calculation of a likelihood function by forward projection, and maximization of the function by Newton's method. Iterative methods have proven efficient for limited data reconstructions, and the intense calculations are no longer considered a problem. All images in this study have been reconstructed with a 3 mm slice thickness and without any filtering post processing.

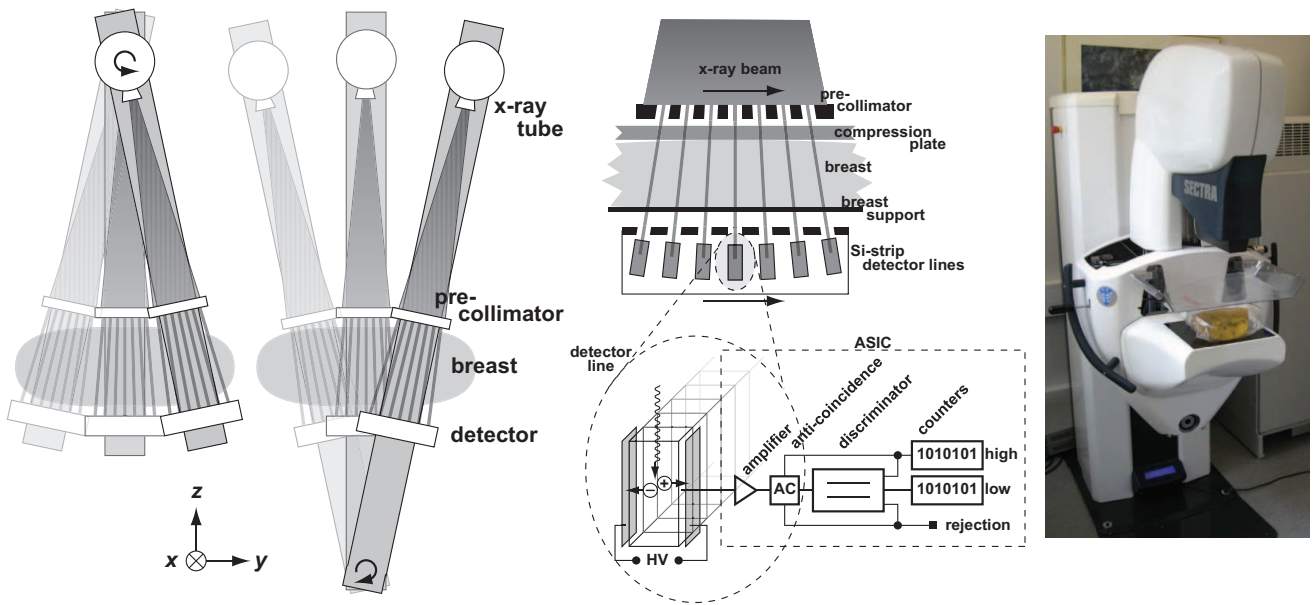

**Figure 1. Left:** In the MDM system, the arm is rotated around the center of the source to acquire a 2D image. In the HighReX system the center of rotation is shifted to a point below the detector so that the source movement describes an arc above the object. Each point in the object is hence viewed from a range of source-detector angles. **Middle:** A close up of the image receptor and the electronics. **Right:** A photo of the prototype HighReX system.

### 2.2.2. The detector

The image receptor in the HighReX system consists of silicon-strip detectors with a strip-pitch of 50 $\mu$m, arranged edge-on for high quantum efficiency despite the low atomic number of silicon (Fig. 1, middle). There are 21 detector lines, which determine the number of projections in the tomosynthesis acquisitions. For each line, there is a corresponding collimator slit in the pre-breast collimator.

A bias voltage is applied over the strip detector (Fig. 1, middle). When a photon interacts with the detector material by the photoelectric effect, charge is released and drifts as electron-hole pairs towards the anode and cathode respectively. Each strip is wire bonded to a preamplifier and shaper, which collect the charge and convert it to a pulse. The pulse height is proportional to the charge and thus to the energy of the impinging photon. Pulses below a certain threshold are regarded as noise and are rejected by a discriminator. All remaining pulses are registered by a counter. A preamplifier with discriminator and counter is referred to as a channel, and all channels are implemented in an application specific integrated circuit (ASIC). Similar detectors and electronics are in use in the Sectra MDM system.[11, 12, 15]

A new feature of the present ASIC is an additional higher threshold of the discriminator with a corresponding additional counter, which is used to sort the detected pulses into two bins according to energy, and thus to allow for DES imaging as described above.

A second novelty is anti-coincidence (AC) logic to remedy the effects of charge sharing. Charge sharing occurs when a photon is absorbed close to the border between two strips and the charge is split between the corresponding channels. Without AC, two photons of lower energy would be detected instead of one high-energy photon, which would increase the noise since the shared photon is double counted (counted twice). Charge sharing would also degrade the spatial resolution since the signal is blurred between two pixels, and the energy resolution since the energy is divided into two parts. The AC logic distinguishes charge-shared events by a simultaneous detection in two adjacent channels, and the event is registered only once in the high-energy bin of the channel with the largest signal. Note that the shared charge needs to be large enough to reach over the low-energy threshold in order to be detected.

### 2.2.3. The DES measurement set-up

As noted above, DES is not yet implemented in the HighReX system, and measurements were therefore performed with a single 128-channel detector element in a set-up similar to a 2D scanned-slit mammography geometry. The effective source size was $400 \times 461$ $\mu$m, and the beam was filtered with aluminum absorption filters. A collimator slit up-stream of the object matched the detector, and the object was scanned across the beam to acquire an image. The lower threshold of the detector was set to discriminate against pulses below approximately 13 keV, and the higher threshold to split the spectrum at approximately 33 keV, corresponding to the K-edge of iodine. To measure the energy spectrum, a CZT compound solid state detector* was inserted in the beam path. It has a near 100% detection efficiency and negligible hole tailing in the considered energy interval.

DES measurements in the same set-up have been presented in a previous study,[7] and here we evaluate some of the results further. A phantom was used, consisting of a PMMA slab with 1-9 mm deep containers filled with iodinated contrast agent.† The iodine was diluted, and the concentration was measured by x-ray absorption to 3.75 mg/ml, which is a realistic concentration for tumor uptake.[16] To hide the iodine containers, the slab was covered with antropomorphic clutter made of olive oil and PMMA. These constituents correspond in absorption difference approximately to adipose and fibroglandular tissue,[6] and the anatomical noise was shown to correspond fairly well with real breast tissue.[7] The total thickness of the phantom was 4.5 cm. A wedge with PMMA-to-oil fractions ranging from 10% to 90% was used to simulate a variety of glandularities for determination of $w$. The average glandular dose (AGD) was calculated by applying normalized glandular dose coefficients[17] to a measured spectrum.

## 2.3. Measurements and simulations

### 2.3.1. Tomosynthesis measurements

Clinical trials were performed with the HighReX system during the fall of 2008 at Capio/St. Göran's Hospital in Stockholm, Sweden. Samples of the acquired images are shown in this study.

To measure the PSF in the $y$-$z$-plane, we used a 50 $\mu$m thin tungsten wire slanted in the plane at 12.3° to the $y$-axis. The 2D PSF, which is pre-sampled in $z$ but not in $y$, was calculated from a single slice in the reconstructed volume with the wire angle as input.[3, 18] The in-plane PSF in $y$ was measured at the waist of the 2D PSF. As noted above, this measurement may give misleading results, but it was nevertheless included to facilitate comparisons to other studies. The in-plane PSF in the $x$-direction is on the contrary a meaningful quantity as it is virtually independent of the $y$-$z$-resolution. It was measured with the same wire, but slanted in the $x$-$y$-plane for over-sampling. The measurement was somewhat complicated by the strong edge enhancement in the $y$-direction; the wire is erroneously enhanced by its' $y$-component. We overcame this problem by using a very slight angle and over sample at well separated points on the wire. From the 1D and 2D PSF's, the MTF's were calculated with 1D and 2D fast Fourier transforms.

---
*Amptek XR-100T-CZT
†Ultravist 370, BayerSchering, Germany

**2.3.2. DES measurements and simulations**

The purpose of the DES measurements was to fully understand previous experimental results, and to provide input to a model of the detector energy sensitivity.

Firstly, the stability of the energy thresholds over time was quantified by recording the high- and low-energy count rates in a flat-field image during a long period of time. Threshold fluctuations have been observed previously,[7] and it is important to verify that the time scale is larger than a typical acquisition time of 10 s.

Secondly, the linearity of the detector was measured as the count rate at an increasing photon-absorption rate. By fitting the function $r = R \cdot \exp(-R\tau)$ to the recorded count rate, the dead time $\tau$ can be calculated.[19] $R$ is the photon rate, which was extrapolated from the linear curve through points at very low count rates. A linearity measurement is important when considering energy resolution in order to avoid pile-up. At count rates that are approaching $1/\tau$, two or more photons may hit the same detector element within one read out, and a single high-energy photon is recorded instead of several photons of lower energy.

Thirdly, the detector model was set up using the MATLAB software package.[‡] Four different effects were taken into account; quantum efficiency, intrinsic energy resolution of silicon, charge sharing and the AC logic, and the spread in threshold levels between different channels.

Quantum efficiency was calculated with published linear absorption coefficients[20] as input. The 500 $\mu$m thick silicon wafer was arranged at an angle of 4°, which yields an effective detector thickness of 7.2 mm. Photo-electric events were detected, whereas scattered events only filtered the beam and no secondary interactions of scattered photons were considered.

The intrinsic energy resolution due to a low number of released charge pairs was modeled as normal distributed with standard deviation $\sigma = \sqrt{E\eta\epsilon}$, where $E$ is the photon energy, $\epsilon = 3.66$ eV is the energy needed to create an electron hole pair in silicon, and $\eta = 0.115$ is the Fano factor for silicon.[19]

Charge-sharing data for a similar detector[15] was used to calculate the fraction of events that share enough energy with the adjacent strip to reach over the low-energy threshold and trigger the AC logic, and the decline in energy of those that do not. We do, however, anticipate some leakage in the logic, i.e. charge-shared events that are not caught by the logic and therefore double counted. Leakage correlates the pixel noise and can therefore be seen as a bent noise power spectrum (NPS) in the detector direction. To estimate the leakage, $M = 512$ $128{\times}128$ regions of interest (ROI's) were acquired from a flat-field image of 45 mm PMMA. The high-energy threshold was maximized to record charge-shared events exclusively in the high-energy bin. The NPS was calculated as the mean of the squared fast Fourier transform of the difference in image signal from the mean in each ROI. If no double counting is present, the NPS at $\nu_x = \nu_x = 0$ (NPS(0)) is the variance of the total integrated pixel values between all ROI's, but it is somewhat larger if there is leakage. The fraction double-counted photons ($\chi$) and the fractional leakage ($\xi$) can therefore be calculated according to

$$\chi = \frac{\text{NPS}(0) - L - H}{3(L+H) - \text{NPS}(0)} \quad \text{and} \quad \xi = \frac{\text{NPS}(0) - L - H}{\text{NPS}(0) - L + H}, \quad (3)$$

where $L$ and $H$ are the average signals in the low- and high-energy bins respectively. A detailed description of the procedure can be found elsewhere.[11]

A spectrum measured with the CZT spectrometer was processed in a model with the three described detector effects. The energy levels of individual channel thresholds were then determined by comparing the measured relative count rate in the high- and low-energy bins with the model. The spread in threshold levels was assumed to be normal distributed, and the standard deviation was calculated.

---

[‡]The MathWorks Inc., Natick, Massachusetts

# 3. RESULTS
## 3.1. Tomosynthesis

Figure 2 shows two different slices from a tomosynthesis breast image acquired during clinical trials with the HighReX system. A group of micro calcifications is clearly seen (left), and a cyst is found further down in the breast (right). These images serve to illustrate the capabilities of the system.

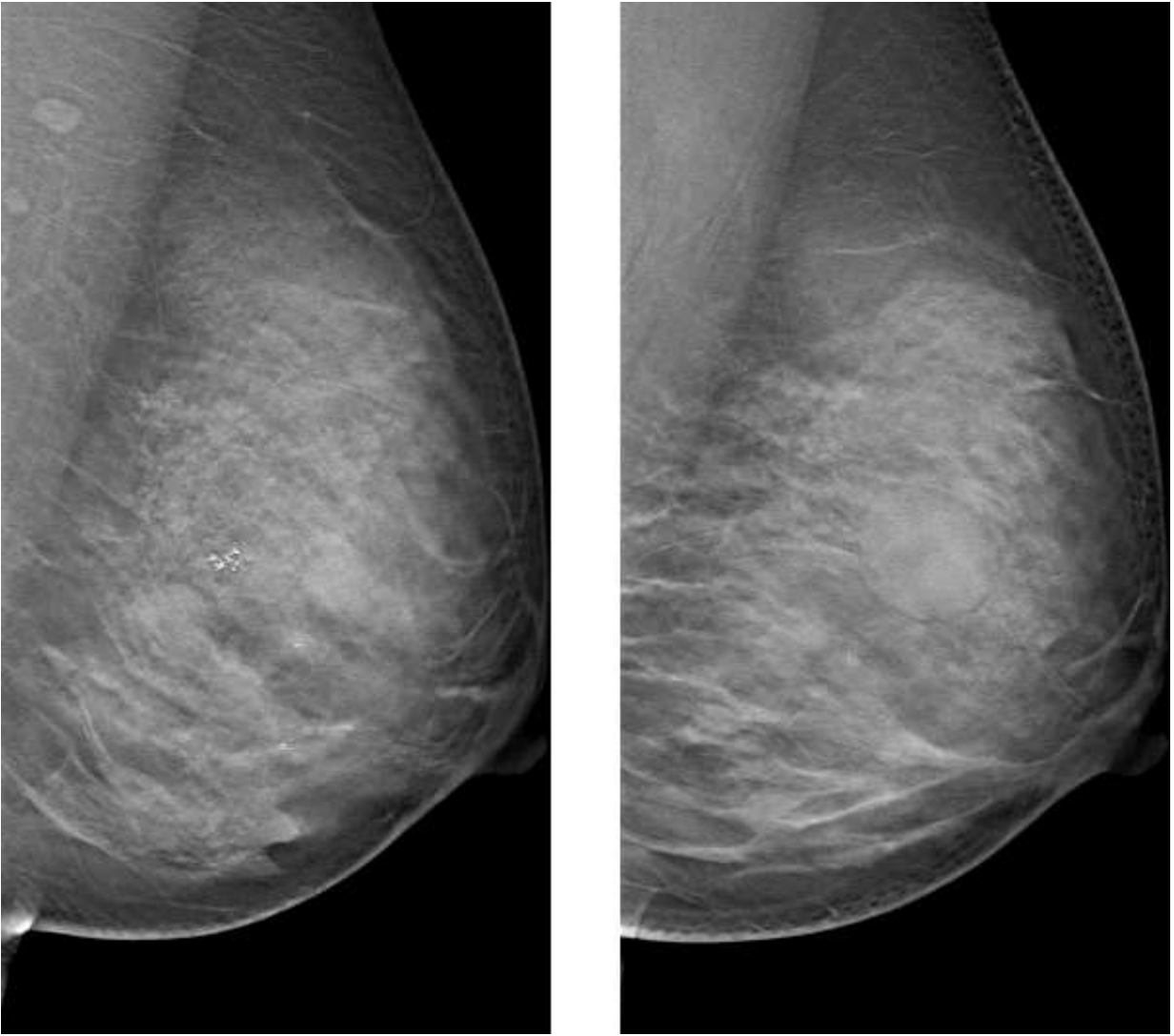

**Figure 2.** Two slices from a tomosynthesis breast image acquired during clinical trials with the HighReX system. **Left:** A group of micro calcifications. **Right:** A cyst is found further down in the breast. Image courtesy of the HighReX project.

Figure 3 shows the in-plane PSF's in $x$ and $y$ with full-widths-at-half-maximum (FWHM's) of 51 and 178 $\mu$m respectively. The dips around the $y$-peak correspond to edge-enhancement. In Fig. 4, the in-plane MTF's are shown together with previously presented results for the MDM system.[13] The agreement is almost perfect in $x$, which is expected, and any discrepancies are easily covered by differences in the measuring methods. In $y$, the tomosynthesis resolution is slightly worse than the projection measurement, but there is edge enhancement, which is visible as a bump around $\nu_y = 1$ mm$^{-1}$.

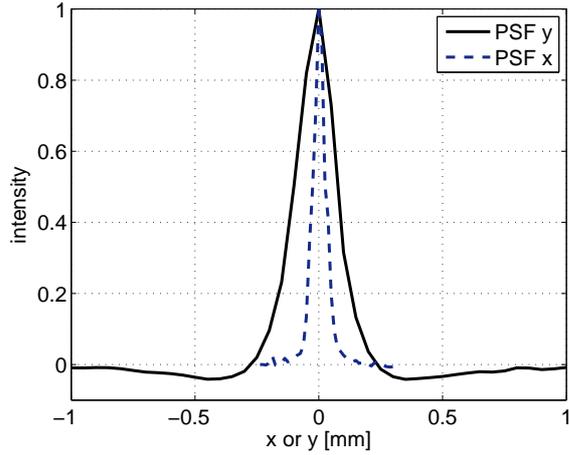

**Figure 3.** The in-plane PSF in $x$ and $y$.

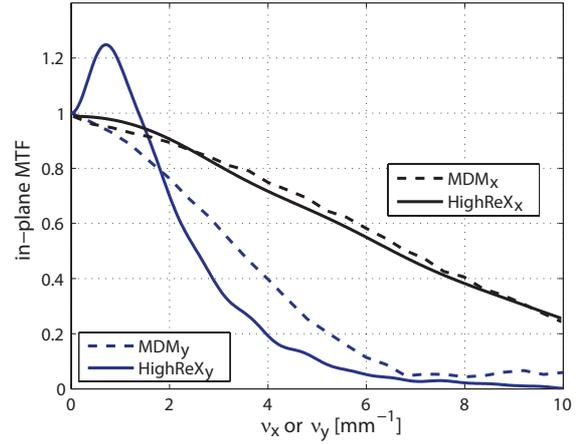

**Figure 4.** The in-plane MTF in $x$ and $y$ compared to published results for the MDM system.

As was described above, the in-plane resolution in $y$ is of limited interest as it does not take the complete PSF into account. Instead, Fig. 5 shows the 2D $y$-$z$ PSF along with the cross section in $z$. The (FWHM) is 2.7 mm, which indicates high anisotropy compared to $y$, and the long tails forecast poor low-frequency resolution. Figure 6 shows the 2D MTF in $\nu_y$ and $\nu_z$, along with projections at low and medium frequencies in both directions. The $y$-resolution at low $\nu_z$ is slightly worse than the in-plane resolution, and the suppression of low $\nu_y$ grows stronger at higher $\nu_z$. In $z$, the MTF at low $\nu_y$ drops quickly, and everything above approximately 0.05 mm$^{-1}$ can be regarded aliasing. At higher $\nu_y$, the $z$-resolution extends 2-4 times longer.

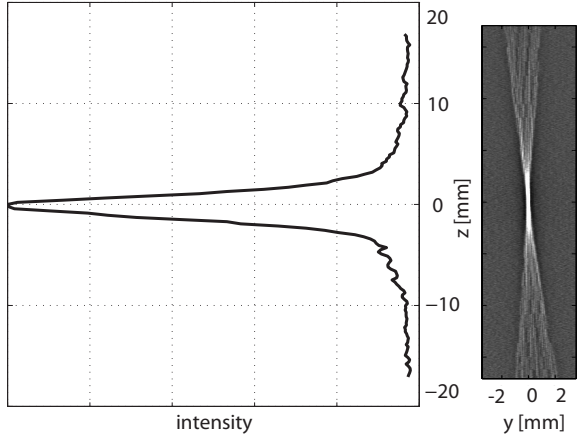

**Figure 5.** The 2D PSF in $y$ and $z$, with the intensity distribution along the $z$ axis.

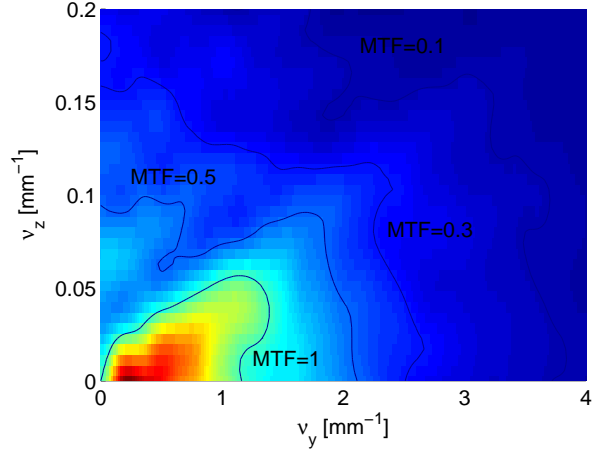

**Figure 6.** The 2D MTF in $y$ and $z$.

### 3.2. Dual-energy subtraction

Figure 8, shows an absorption image of the oil/PMMA phantom at an AGD of 0.5 mGy, and the DES image calculated with $w = 0.72$. None of the containers is distinguishable in the absorption image, whereas at least five are visible in the subtracted image.

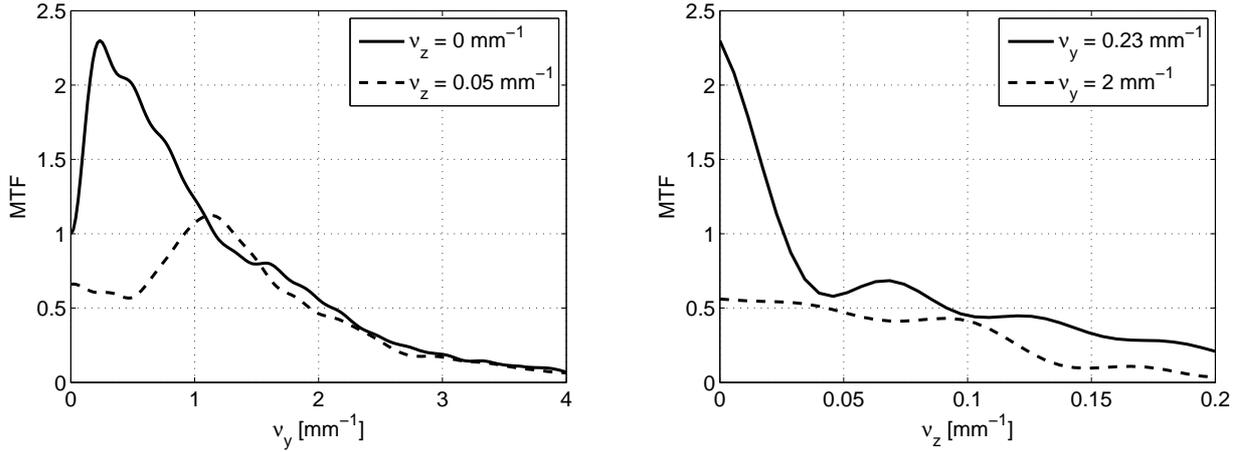

**Figure 7.** Projections of the 2D MTF along the $\nu_y$-axis (left), and $\nu_z$-axis (right).

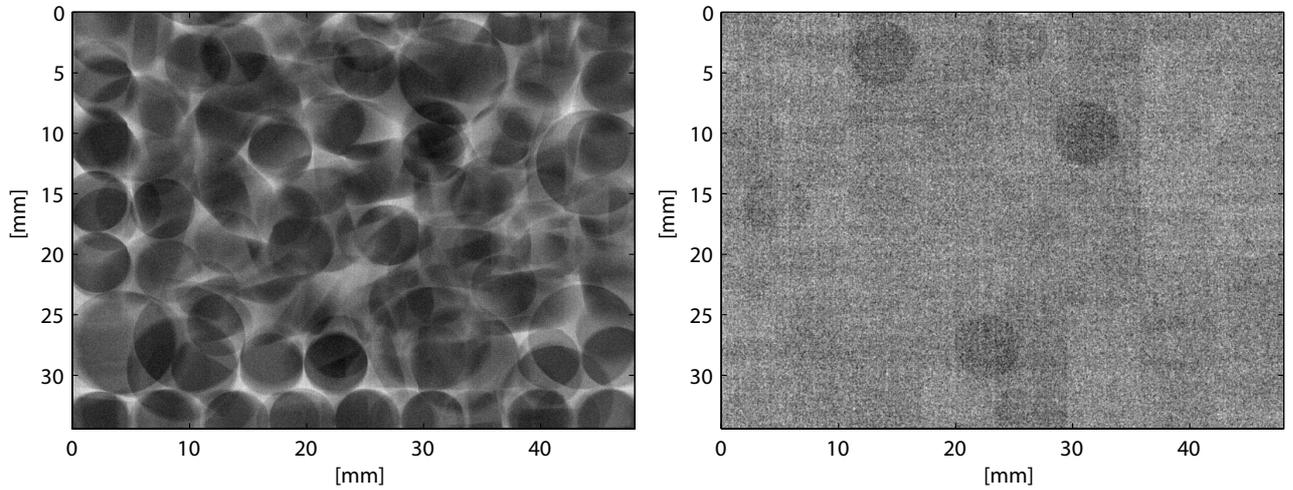

**Figure 8. Left:** An absorption image of the anatomical clutter phantom with 1-9 mm iodine containers. **Right:** The corresponding DES image calculated with $w = 0.72$. The five deepest holes are visible.

The DES image in Fig. 8 exhibits contrast variations over the approximately two-hour acquisition. Such variations are further illustrated in Fig. 9, where a long-term decline of 0.36% per hour is seen along with short-term fluctuations. All variations are likely due to a drifting high-energy threshold, since the DES signal is independent of tube output and the absorption signal, also shown in the figure, is virtually constant. The statistical errors of the measurements were comparably small. During the typical time scale of a clinical acquisition (10 s) the maximum variation in the DES signal is 0.30%, which is negligible for most purposes.

Figure 10 shows the photon rate and detected count rate as a function of x-ray tube current times the exposed detector width. The dead time was calculated to $\tau = 247$ ns, corresponding to a read-out rate of 4.05 MHz. For all measurements presented here, the count-rate was approximately 30 kHz, thus more than an order of magnitude lower, and pile-up should be negligible.

The NPS was measured and used to find the fractional leakage, $\xi = 13\%$, and the fraction double-counted photons, $\chi = 1.0\%$, according to Eq. 3. These values are in correspondence with previous results for similar detectors.[11] The mean of the energy thresholds was determined to 30 keV with a standard deviation of ±1.3 keV.

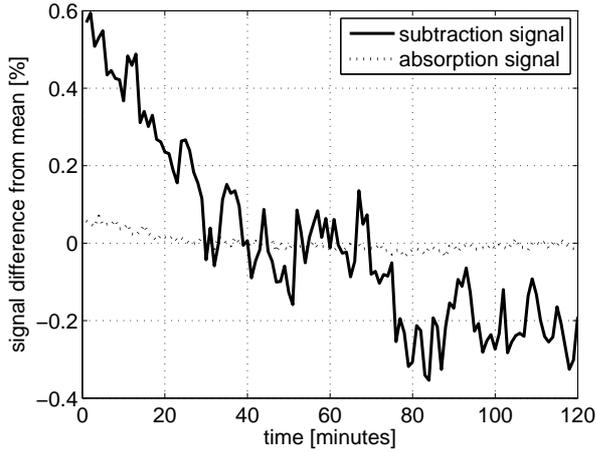

**Figure 9.** Stability of the detector over time. The absorption signal is virtually constant, whereas the DES signal exhibits fluctuations.

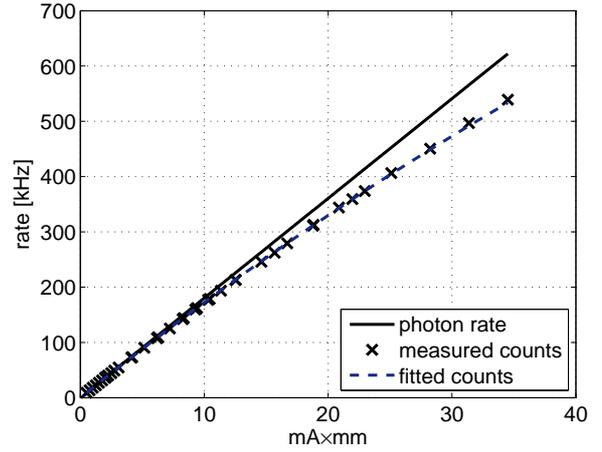

**Figure 10.** The linearity of the detector. The decline in detected count rate due to pile-up at higher photon rates is clearly seen.

Figure 12 finally shows the predicted DES signal difference ($\Delta I_{\text{DES}}$) for the PMMA/oil phantom compared to measured values. Large fluctuations in the measured $\Delta I_{\text{DES}}$ are seen, which should again be due to time variations since the statistical errors are small. A least-square fit of the measurements (with the signal at 9 mm considered an outlier), however, is in almost perfect agreement with the model prediction for the experimental detector. The model predicts that an optimized detector with no threshold spread, a split frequency of 33 keV, and perfect AC logic that discards charge-shared photons instead of directing them to the high-energy bin, enables an improvement in $I_{\text{DES}}$ of 38%.

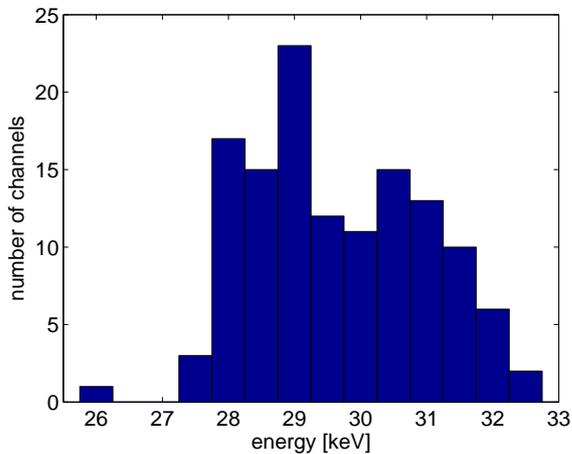

**Figure 11.** The spread of the high-energy thresholds. The mean is 30 keV with a standard deviation of ±1.3 keV.

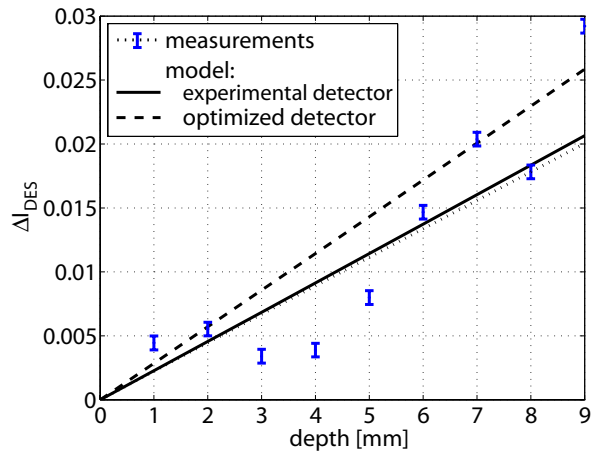

**Figure 12.** The DES signal difference ($\Delta I_{\text{DES}}$) as a function of depth of the iodine containers. Measurements are presented as error bars (one standard deviation statistical uncertainty) and a least-square fit. These are compared to model predictions.

## 4. DISCUSSION

### 4.1. Tomosynthesis

The slanted-wire PSF measurement scheme used in this study is associated with some limitations. Firstly, the PSF is not pre-sampled in $y$, and the wire width is similar to the pixel size. It may be possible to measure a pre-sampled PSF with a wire that is slanted also in $y$, but that would add uncertainty because of the simultaneous angle in $z$. As for the wire thickness, thinner wires were found not to provide enough contrast for a reliable measurement. Secondly, the PSF is not shift invariant throughout the volume. It differs in $z$ due to different magnification factors, and possibly also in $y$ due to different measurement angles.[21] For these reasons, the MTF presented here should be regarded indicative and not exact.

The MTF is not sufficient to fully characterize the system, but should be compared with a noise measurement, ideally in terms of an NPS to form the detective quantum efficiency (DQE) of the system. The NPS has been measured in tomosynthesis in the past, and we are currently preparing a measurement for our system. Future work also include setting up a model for the full system, to be combined with the presented detector model.

### 4.2. DES imaging

The detector was not perfectly optimized for DES imaging with a fairly large spread in the energy thresholds and a too low mean value. Moreover, the current anti-coincidence scheme degrades the energy resolution since shared photons are put in the high-energy bin, regardless of energy. A better alternative would be to separate charge shared photons into a third bin, not using them for DES imaging but still saving the information for transmission images. This study showed that an optimized detector with respect to these deficiencies would improve the subtraction signal substantially, and it is likely that it would perform better also for reducing the anatomical noise. Pile-up was not included in the present detector model, but it is a prioritized upgrade.

## 5. CONCLUSIONS

We have experimentally characterized a prototype for breast tomosynthesis in terms of the MTF. It is confirmed that $z$-resolution is achieved at the cost of $y$-resolution, whereas $x$-resolution is unaffected compared to projection imaging. The full impact of the MTF on detecteability is yet to be determined, but clinical images indicate that the trade-off between $y$ and $z$ is justifiable as depth resolution helps increase the conspicuity of breast lesions. The slanted wire approach that was used to measure the MTF gives expected and reasonable results, but the method is associated with some limitations that need to be taken into account.

We further conclude that the new detector work satisfactorily with a high-energy threshold drift that is slow compared to clinical acquisition times, a short dead time, and an acceptable leakage in the AC logic. The proposed model was successfully verified with experimental results, and predicted that the DES signal difference can be improved by 38% with a few relatively straightforward improvements.

## ACKNOWLEDGMENTS

The authors wish to thank Alexander Chuntonov for setting up the silicon-strip detector. Figure 2 is image courtesy of the HighReX project.

## REFERENCES

1. J. T. 3rd Dobbins and D. J. Godfrey, "Digital x-ray tomosynthesis: current state of the art and clinical potential," *Phys. Med. Biol* **48**(19), pp. 65–106, 2003.
2. G. Wu, J. G. Mainprize, J. M. Boone, and M. J. Yaffe, "Evaluation of scatter effects on image quality for breast tomosynthesis," in *Proc. SPIE, Physics of Medical Imaging*, M. J. Flynn and J. Hsieh, eds., **6510**, 2007.
3. Y. H. Hu, B. Zhao, and W. Zhao, "Image artifacts in digital breast tomosynthesis: Investigation of the effects of system geometry and reconstruction parameters using a linear system approach," *Med. Phys.* **35**(12), pp. 5242–5252, 2008.

4. P. Johns, D. Drost, M. Yaffe, and A. Fenster, "Dual-energy mammography: initial experimental results," *Med. Phys.* **12**, pp. 297–304, 1985.
5. J. Lewin, P. Isaacs, V. Vance, and F. Larke, "Dual-energy contrast-enhanced digital subtraction mammography: Feasibility," *Radiology* **229**, pp. 261–268, 2003.
6. H. Bornefalk and M. Lundqvist, "Dual-energy imaging using a photon counting detector with electronic spectrum-splitting," in *Proc. SPIE, Physics of Medical Imaging*, M. Flynn and J. Hsieh, eds., **6142**, 2006.
7. E. Fredenberg, B. Cederström, M. Lundqvist, C. Ribbing, M. Åslund, F. Diekmann, R. Nishikawa, and M. Danielsson, "Contrast-enhanced dual-energy subtraction imaging using electronic spectrum-splitting and multi-prism x-ray lenses," in *Proc. SPIE, Physics of Medical Imaging*, J. Hsieh and E. Samei, eds., **6913**, 2008.
8. "The HighReX Project (High Resolution X-ray imaging)." online: http://www.highrex.eu.
9. M. Åslund, B. Cederström, M. Lundqvist, and M. Danielsson, "Scatter rejection in multi-slit digital mammography," *Med. Phys.* **33**, pp. 933–940, 2006.
10. R. N. Cahn, B. Cederstrom, M. Danielsson, A. Hall, M. Lundqvist, and D. Nygren, "Detective quantum efficiency dependence on x-ray energy weighting in mammography," *Med. Phys.* **26**(12), pp. 2680–3, 1999.
11. M. Lundqvist, *Silicon strip detectors for scanned multi-slit x-ray imaging*. PhD thesis, Royal Institute of Technology (KTH), Stockholm, 2003.
12. E. Beuville, B. Cederström, M. Danielsson, L. Luo, D. Nygren, E. Oltman, and J. Vestlund, "An application specific integrated circuit and data acquisition system for digital x-ray imaging," *Nucl. Instr. Meth. A* **406**, pp. 337–342, 1998.
13. M. Åslund, B. Cederström, M. Lundqvist, and M. Danielsson, "Physical characterization of a scanning photon counting digital mammography system based on Si-strip detectors," *Med. Phys.* **34**(6), pp. 1918–1925, 2007.
14. K. Lange and J. A. Fessler, "Globally convergent algorithms for maximum a posteriori transmission tomography," *IEEE Transactions on Image Processing* **4**(10), pp. 1430–1438, 2005.
15. M. Lundqvist, B. Cederström, V. Chmill, M. Danielsson, and D. Nygren, "Computer simulations and performance measurements on a silicon strip detector for edge-on imaging," *IEEE Trans. Nucl. Science* **47**(4), pp. 1487–1492, 2000.
16. A. Teifke, F. Schweden, H. Cagil, H. Kauczor, W. Mohr, and M. Thelen, "Spiral-Computertomographie der Mamma," *Rofo* **161**(6), pp. 495–500, 1994.
17. J. Boone, "Glandular breast dose for monoenergetic and high-energy x-ray beams: Monte carlo assessment," *Radiology* **203**, pp. 23–37, 1999.
18. M. J. Flynn, R. McGee, and J. Blechinger, "Spatial resolution of x-ray tomosynthesis in relation to computed tomography for coronal/sagittal images of the knee," in *Proc. SPIE, Physics of Medical Imaging*, M. Flynn and J. Hsieh, eds., **6510**, 2007.
19. W. R. Leo, *Techniques for nuclear and particle physics experiments*, Springer-Verlag, 1987.
20. M. Berger, J. Hubbell, S. Seltzer, J.S., Coursey, and D. Zucker, "XCOM: Photon Cross Section Database." online: http://physics.nist.gov/xcom. National Institute of Standards and Technology, Gaithersburg, MD, 2005.
21. K. Israni, G. Avinash, and B. Li, "Point Spread Function based classification of regions for Linear Digital Tomosynthesis," in *Proc. SPIE, Physics of Medical Imaging*, M. Flynn and J. Hsieh, eds., **6510**, 2007.